\documentclass[12pt]{article}
\usepackage{latexsym}

\newcommand{\beq}{\begin{equation}}
\newcommand{\eeq}[1]{\label{#1}\end{equation}}
\newcommand{\bea}{\begin{eqnarray}}
\newcommand{\eea}[1]{\label{#1}\end{eqnarray}}

\begin{document}
\begin{titlepage}
\hfill NYU-TH/02/02/01 hep-th/0203014
\vspace{20pt}

\begin{center}
{\large\bf{FULLY COVARIANT VAN DAM-VELTMAN-ZAKHAROV DISCONTINUITY, 
AND ABSENCE THEREOF}}
\end{center}

\vspace{6pt}

\begin{center}
{\large M. Porrati} \vspace{20pt}

{\em Department of Physics, NYU, 4 Washington Pl, New York NY 10003}

\end{center}

\vspace{12pt}

\begin{center}
\textbf{Abstract }
\end{center}
\begin{quotation}\noindent
In both old and recent literature, it has been argued that the celebrated 
van Dam-Veltman-Zakharov (vDVZ) discontinuity of massive gravity is an
artifact due to linearization of the true equations of motion.
In this letter, we investigate that claim.
First, we exhibit an explicit --albeit somewhat arbitrary-- fully 
covariant set of equations of motion that, upon linearization, 
reduce to the standard Pauli-Fierz equations. We show that the vDVZ 
discontinuity still persists in that non-linear, covariant theory.
Then, we restrict our attention to a particular system that consistently
incorporates massive gravity: the Dvali-Gabadadze-Porrati (DGP) model.
DGP is fully covariant and does not share the arbitrariness and 
imperfections of our previous covariantization, and its linearization exhibits
a vDVZ discontinuity. Nevertheless, we explicitly
show that the discontinuity does disappear in the fully covariant theory, and
we explain the reason for this phenomenon. 
\end{quotation}
\vfill
\end{titlepage}
\section{Introduction} 
In a famous paper, van Dam and Veltman~\cite{vdv} (see also Zakharov~\cite{z})
studied a massive spin-2 field that couples to matter as the graviton, namely
as $h^{\mu\nu}T_{\mu\nu}$ ($T_{\mu\nu}$ is the conserved stress-energy 
tensor). They showed that, at distances much smaller than the Compton 
wavelength of the massive graviton, one recovers Newton's law by an 
appropriate choice of the spin-2 coupling constant. On the other hand, in the
small-mass limit, the bending angle of light by a massive body approaches 
$3/4$ of the Einstein result. This is the vDVZ discontinuity. 
A physical explanation of this phenomenon is that a massive spin-2 
field carries 5 polarizations, whereas a massless one carries only two.
In the limit $m\rightarrow 0$, therefore, 
a massive spin-2 field decomposes into 
massless fields of spin 2, 1, and 0. The spin-0 field couples to the trace of
the stress-energy tensor, so that in the limit $m\rightarrow 0$ one does not 
recover Einstein's gravity but rather a scalar-tensor theory.

This result seems to rule out any modification of Einstein's gravity in 
which the principle of equivalence still holds, but the graviton acquires a 
mass, no matter how tiny. 

In the presence of a negative cosmological constant $\Lambda$, 
on an Anti de Sitter background, the one-graviton
amplitude between conserved sources is continuous in the limit
$m^2/\Lambda\rightarrow 0$~\cite{p,kmp}, so that one cannot rule a massive
graviton with a Compton wavelength of the order of the Hubble scale. In 
refs.~\cite{kr,p2}, it was shown from various viewpoints 
that the AdS graviton may indeed become 
massive, when standard gravity is coupled to conformal matter. 

On a de Sitter background, a massive spin-2 field is unitary only if
$m^2\geq 2\Lambda/3$~\cite{h}. 

All of this makes perfect sense, yet, the very fact that experiments 
at a scale of roughly an astronomical unit can tell that the mass of the 
graviton is smaller than the inverse Hubble radius is baffling to some.
After all, the latter scale is $10^{16}$ times smaller than the former!

In fact, several old~\cite{v} and recent~\cite{nyu1,nyu2,nyu3} 
papers have claimed or argued that the vDVZ 
discontinuity is an artifact of the Pauli-Fierz Lagrangian, i.e. of the 
linearization of the true, covariant, non-linear equations of massive gravity.

Most of the renewed attempts to go beyond the linear approximation to massive 
gravity have exploited the DGP model~\cite{dgp}, which has both a 
massive-graviton spectrum and four-dimensional general covariance.
 
In this letter, we would like to study the existence of 
a general covariant vDVZ discontinuity from two points of view.

First of all, in Section 2,
we exhibit explicitly covariant, fully nonlinear equations
of motion for a massive spin-2 field coupled to matter that, after 
linearization, reduce to the Pauli-Fierz system studied in~\cite{vdv,z}.
We show that the discontinuity found in the linearized equations 
persists at the non-linear, fully covariant level by finding a covariant
constraint not present in standard, massless gravity. 
This definitely settles in the affirmative the question of 
whether a general-covariant extension of the vDVZ result exists.

In Section 3, we study the DGP model, for two reasons. 
The first is that it is a
promising candidate for a brane-world realization of gravity, which may even
shed light on the cosmological constant problem. The second is that the 
covariantization studied in the first part of the paper is far from being
unquestionable. Besides being somewhat arbitrary, so that it does not rule 
out the possibility of other discontinuity-free covariantizations, it
is also non-local. Non-locality signals the presence of other light, possibly
unphysical degrees of freedom coupled to ordinary matter (negative-norm 
ghosts, for instance).
The DGP model, instead, is a consistent model that exhibits a vDVZ 
discontinuity at linear order. We show that there, 
as argued in~\cite{nyu1,nyu2,nyu3}, the discontinuity {\em does} 
indeed disappear when the DGP is studied beyond its linear
approximation. To prove that, we relate the breakdown of the linear
approximation to the fact that the brane can bend in the fifth dimension, so
that its induced curvature may be large even when the source on the brane is
weak.

\section{Covariantization of the Pauli-Fierz Action}
A long time ago, Pauli and Fierz~\cite{pf} found a local, 
covariant  action describing a free, massive spin-2 field. The action is 
unique  up to field redefinitions and it reads
\beq
S=S_L[h_{\mu\nu}] + \int d^4x \left[ {M^2\over 64\pi G}
(h_{\mu\nu}^2 -h^2) -{1\over 2}h_{\mu\nu}T^{\mu\nu}\right].
\eeq{1}
Here, $T_{\mu\nu}$ is an external source, that we assume to be conserved 
and identify with the stress-energy tensor of matter.
$S_L[h_{\mu\nu}]$ is the Einstein action expanded to quadratic order in the
metric fluctuations around flat space:
\bea
S_L[h_{\mu\nu}]&=& 
{1\over 64\pi G}\int d^4x h^{\mu\nu} L_{\mu\nu, \rho\sigma}h^{\rho\sigma},
\nonumber \\
S_E[\eta_{\mu\nu}+h_{\mu\nu}]&=&{1\over 16 \pi G} \int d^4x 
\sqrt{-\det( \eta +h)}R(\eta+h)=S_L[h_{\mu\nu}] +O(h^3).
\eea{2}
At $M=0$, the action in Eq.~(\ref{1}) is obviously invariant under 
linearized diffeomorphisms
\beq
h_{\mu\nu} \rightarrow h_{\mu\nu} + \partial_\mu \xi_\nu +
\partial_\nu \xi_\mu,
\eeq{3}
but the mass term explicitly breaks this invariance.
A gauge-invariant form of the Pauli-Fierz action is achieved by using the 
St\"uckelberg mechanism, i.e. by adding a vector field that transforms linearly
under local diffeomorphisms:
\beq
A_\mu \rightarrow A_\mu -\xi_\mu.
\eeq{4}
Substituting $h_{\mu\nu}\rightarrow h_{\mu\nu} + 2\partial_{(\mu} 
A_{\nu)}$ in Eq.~(\ref{1}) we find the manifestly gauge-invariant 
``St\"uckelberg-Pauli-Fierz'' (SPF) action
\beq
S_{SPF}[h]= S_L[h] + \int d^4x \left\{ {M^2\over 64\pi G}
[(h_{\mu\nu} + 2\partial_{(\mu} A_{\nu)})^2 -(h + 2\partial\cdot A)^2] -
{1\over 2}h_{\mu\nu}T^{\mu\nu}\right\}.
\eeq{5}
The gauge-invariant equations of motion are
\bea
L_{\mu\nu,\rho\sigma}h^{\rho\sigma} + M^2 [h_{\mu\nu} + 
2\partial_{(\mu} A_{\nu)} -\eta_{\mu\nu}(h + 2\partial\cdot A)] &=& 
16\pi G T_{\mu\nu}, \label{6} \\
\partial^\nu F_{\nu\mu} + J_\mu=0, \qquad J_\mu &=& \partial^\nu h_{\mu\nu} - 
\partial_\mu h. 
\eea{7}
Of course, $F_{\mu\nu}= \partial_\mu A_\nu -\partial_\nu A_\mu$.
Notice that the Pauli-Fierz mass term is precisely the combination that
gives a gauge-invariant equation of motion for $A_\mu$. Equation~(\ref{7})
is easily solved by
\beq
A_\mu= -\Box^{-1} J_\mu + \partial_\mu \phi.
\eeq{8}
$\phi$ is an arbitrary function since Eq.~(\ref{7}) is invariant under the 
gauge transformation $A_\mu \rightarrow A_\mu +\partial_\mu \chi$.
We can then select a particular solution to Eq.~(\ref{7}) by choosing 
$\phi= -\Box^{-1} h $:
\beq
A_{\mu}= -\Box^{-1} I_\mu, \qquad 
I=\partial^\nu h_{\mu\nu} - {1\over 2}\partial_\mu h.
\eeq{9}
Substituting this $A_\mu$ into Eq.~(\ref{6}) we arrive at a 
particularly interesting form of the equations of motion:
\beq
L_{\mu\nu,\rho\sigma}h^{\rho\sigma} + M^2 [h_{\mu\nu} +h_{\mu\nu} + 
-2\partial_{(\mu} \Box^{-1}I_{\nu)} -\eta_{\mu\nu}(h - 2\Box^{-1}
\partial\cdot I)]= 16\pi G T_{\mu\nu}.
\eeq{10}
Recalling the definition of $I_\mu$, and noticing that 
$L_{\mu\nu,\rho\sigma}h^{\rho\sigma}$ is by construction 
proportional to the linearized Einstein tensor,
$L_{\mu\nu,\rho\sigma}h^{\rho\sigma}=2G^L_{\mu\nu}=
2R^L_{\mu\nu}-\eta_{\mu\nu}R^L$, we can be recast Eq.~(\ref{10}) into 
the suggestive form
\beq
G^L_{\mu\nu} -M^2 \Box^{-1}(R^L_{\mu\nu}-\eta_{\mu\nu}R^L)=8\pi G T_{\mu\nu}.
\eeq{11}  

It is now obvious how to promote the Pauli-Fierz equations 
into a fully covariant form. First, we notice that any symmetric tensor 
$S_{\mu\nu}$ can be decomposed as $S_{\mu\nu}=S^T_{\mu\nu} + 
D_{(\mu}S_{\nu)}$, $D^\mu S^T_{\mu\nu}=0$. Then, we replace all linearized 
tensors in Eq.~(\ref{11}) with their exact form
\beq
G_{\mu\nu} -M^2 \left(\Box^{-1}G_{\mu\nu}\right)^T + {1\over 2} M^2 
g_{\mu\nu}\Box^{-1} R
=8\pi G T_{\mu\nu},
\eeq{12}
where $T_{\mu\nu}$ obeys the {\em covariant} conservation equation
$D^\mu T_{\mu\nu}=0$. 
We can also find the covariant form of the vDVZ discontinuity.
By taking the double divergence of Eq.~(\ref{12}), we get a new constraint
on the metric, not present in Einstein's gravity:
\beq
D^\mu D^\nu \left[G_{\mu\nu} -M^2 \left(\Box^{-1}G_{\mu\nu}\right)^T + 
{1\over 2}M^2 g_{\mu\nu}\Box^{-1}R\right]=-{M^2\over 2} R=0.
\eeq{13}

Clearly this constraint, implying that the scalar curvature is zero 
everywhere, cannot be satisfied by a metric obeying Einstein's equations up
to a small deformation $o(M)$. Notice also that we would have missed 
the existence of the discontinuity if we only studied the Einstein
vacuum equations, $R_{\mu\nu}=0$. In the covariantization studied here, 
the discontinuity appears only when 
comparing Eq.~(\ref{13}) with the Einstein equations {\em in matter}, where
$R=-8\pi G T\neq 0$.

Eq.~(\ref{12}) is fully covariant and it reduces to the PFS equations to 
linear order, but it is far from satisfactory. 
The first 
problem is that it is by
no means the only covariantization of Eq.~(\ref{11}), so that we cannot 
exclude {\em a priori} that other covariantizations exist, in which the
vDVZ discontinuity disappears.
Secondly, Eq.~(\ref{12}) cannot be derived from a covariant action, 
since if that were the case its covariant divergence would automatically 
vanish, instead of giving the constraint Eq.~(\ref{13}). One could hope that
a ``good'' covariantization, where the divergence of the equations of motions 
vanishes identically, may also cure the discontinuity.
 
 A third, more serious problem, is that
Eq.~(\ref{12}) is nonlocal and it may, therefore, describe the propagation
of other light, possibly unphysical degrees of freedom.

We address the first and third problems in the next Section, 
when discussing a consistent embedding of massive gravity into a ghost-free 
theory: the DGP model. 

The second problem is addressed here, by showing that another covariantization
of Eq.~(\ref{11}) exists, with the desired property that
the covariant divergence vanishes identically, but in which the
vDVZ discontinuity is still present.

First of all, recall that Eq.~(\ref{8}) depends on an arbitrary scalar 
function. We can then write, generically,
\beq
A_\mu= -\Box^{-1} I_\mu + \partial_\mu \varphi.
\eeq{14}
We can also introduce another scalar, $\psi$, and 
redefine the linearized metric as
\beq
h_{\mu\nu}\rightarrow h_{\mu\nu} + \eta_{\mu\nu}\psi. 
\eeq{15}
This redefinition changes the (linearized) Einstein tensor and the scalar 
curvature as follows
\beq
G^L_{\mu\nu}\rightarrow G^L_{\mu\nu} + \eta_{\mu\nu}\Box \psi - \partial_\mu
\partial_\nu \psi, \qquad R^L\rightarrow R^L -3\Box\psi.
\eeq{16}
Thanks to Eqs.~(\ref{14},\ref{16}) we can re-write the PFS equations as
\bea
G^L_{\mu\nu} -M^2 \Box^{-1}\left[G^L_{\mu\nu}-{1\over 2}
\eta_{\mu\nu}\left(R^L -3 \Box \psi\right)\right]+ &&\nonumber \\
(1- M^2 \Box^{-1})(\eta_{\mu\nu}\Box \psi - \partial_\mu\partial_\nu \psi) 
- M^2(\eta_{\mu\nu}\Box \varphi - \partial_\mu\partial_\nu \varphi)
&=& 8\pi G T_{\mu\nu}.
\eea{17}
This equation can be simplified by setting $ 3 \Box \psi=R^L$,
and $(1- M^2 \Box^{-1})\psi=M^2\varphi$:
\beq
(1-M^2\Box^{-1}) G^L_{\mu\nu}=8\pi G T_{\mu\nu}.
\eeq{18}
By taking the trace of Eq.~(\ref{18}), we find $(M^2\Box^{-1}-1)R^L=8\pi GT$, 
so that we can re-write the equation that defines $\psi$ in a more instructive 
form
\beq
(\Box -M^2)\psi =-{8\pi\over 3} G T.
\eeq{19}
The covariantization of Eqs.~(\ref{18},\ref{19}) is now obvious
\beq
G_{\mu\nu}-M^2\left(\Box^{-1} G_{\mu\nu}\right)^T=8\pi G T_{\mu\nu}, \qquad 
(\Box -M^2)\psi =-{8\pi\over 3} G T.
\eeq{20}
Notice that the divergence of the tensor equation is automatically zero
thanks to the covariant conservation of the stress-energy tensor and a
standard Bianchi identity of general relativity. Notice also that the vDVZ
discontinuity is still present, as Eqs.~(\ref{20}) describe a scalar-tensor
theory, in which the massive scalar $\psi$ couples with gravitational strength
to the trace of the stress-energy tensor.
\section{Absence of vDVZ Discontinuity in the DGP Model}
The results of the previous Section seem to indicate that even a consistent 
theory of massive gravity may suffer 
from a vDVZ discontinuity besides the
linear order. Nevertheless, we will show that this is not the case in the
DGP model, as already argued in~\cite{nyu1,nyu2,nyu3} (see also~\cite{nyu4}). 

The DGP model describes a 4-d brane moving in a 5-d space with vanishing 
cosmological constant. 
In five dimensions, the Einstein action is
\beq
S_5=\int_\Sigma d^5 x {1\over 16\pi \hat{G}}\sqrt{-\hat{g}} \hat{R}(\hat{g})+
S_{GH}.
\eeq{21}
Here, hatted quantities are five-dimensional, while un-hatted ones are 
four-dimensional. The integral is performed over a space $\Sigma$ 
that is, topologically, the direct product of the real half-line $R^+$ and
the 4-d Minkowski space $M_4$. We parametrize this space with 
four-dimensional coordinates $x^\mu$, $\mu=0,1,2,3$ and a fifth 
coordinate $y\equiv x^4$. $S_{GH}$ is the Gibbons-Hawking boundary 
term~\cite{gh}, whose explicit form we will not need.

The model is specified by the Einstein equations inside $\Sigma$ and by 
boundary conditions at the brane's position, i.e. at the $\Sigma$ boundary
$\partial \Sigma=M_4$:
\beq
{1\over 16\pi \hat{G}} K_{\mu\nu}\equiv-{1\over \sqrt{-g}}
{\delta S_5 [g] \over \delta g^{\mu\nu}}= 
{1\over 16\pi G} 
\left( R_{\mu\nu}-{1\over 2} g_{\mu\nu} R \right) -{1\over 2} T_{\mu\nu}.
\eeq{22}
As evident from this equation, 
the brane has a 4-d nonzero Newton's constant $G$. 
$T_{\mu\nu}$ is the stress-energy tensor of the matter living on the brane. 
The 4-d cosmological constant is assumed to be negligible. This corresponds
to a limit in which the brane is almost tensionless, and possesses only a 
bending energy, proportional to the extrinsic
curvature $K_{\mu\nu}$.
It is convenient to work in Gaussian coordinates where the brane is located 
at $y=0$, and where 
\beq
g_{\mu 4}(x,y)|_{y=0}=0.
\eeq{22a} 
In these coordinates, 
\beq
K_{\mu\nu}={1\over 2} \sqrt{g_{44}}(\dot{g}_{\mu\nu} - g_{\mu\nu} 
g^{\alpha\beta}\dot{g}_{\alpha\beta}).
\eeq{23}
The dot denotes the derivative w.r.t. $y$.
The linearization of Eq.~(\ref{22}) has been given in~\cite{nyu1,nyu3,gr}. 
It is most conveniently performed in the 5-d harmonic gauge:
\beq
\partial_a h^a_b -{1\over 2} \partial_b h=0 , \qquad \hat{g}_{ab}=\eta_{ab} +
h_{ab}, \qquad a,b=0,..,4.
\eeq{24}
This gauge choice is compatible with Eqs.~(\ref{22a}); indeed, it is 
compatible with setting $g_{\mu 4}=0$ everywhere in $\Sigma$.
After this last gauge choice, the linearized equations assume the simple form
\bea
\Box h_{ab}(z) + \ddot{h}_{ab}(z)&=&0, \qquad h_{\mu 4}(z)=0 ,\qquad z\in 
\Sigma, \label{25}\\ 
{1\over L} [\dot{h}_{\mu\nu}(x)-\eta_{\mu\nu}\dot{h}(x)]&=&
\Box h_{\mu\nu}(x) -\partial_\mu \partial_\nu h(x) +16\pi G T_{\mu\nu}(x),
\qquad x\in \partial\Sigma.
\eea{26}
The ratio $L\equiv \hat{G}/G$ plays a fundamental role in the DGP model, since
it is the transition length beyond which 4-d gravity turns into 5-d gravity.

Eqs.~(\ref{25},\ref{26}) are easily solved by Fourier transforming the 
4-d coordinates $x^\mu$~\cite{nyu1}
\bea
\tilde{h}_{\mu\nu}(p)&=& \tilde{\bar{h}}_{\mu\nu}(p) + p_\mu p_\nu
{16\pi GL\over p^3 + p^2/L} \tilde{T}(p)\exp(-py),
\label{26a} \\
\tilde{\bar{h}}_{\mu\nu}(p)&=&  {16\pi G\over p^2 + p/L} \left[
\tilde{T}_{\mu\nu}(p) -{1\over 3} \eta_{\mu\nu} \tilde{T}(p) \right]\exp(-py).
\eea{27} 
These equations contain a term proportional to $L$, that diverges in the 
decoupling limit $L\rightarrow \infty$. To linear order, this divergence is an
artifact of our gauge choice, in which the brane lies at $y=0$.       
It can be canceled by transforming into new coordinates, $\bar{y}$, 
$\bar{x}^\mu$, in which the brane lies at $\bar{y}=\epsilon_4(x,0)$.
\bea
\bar{x}^\mu &=& x^\mu + \epsilon^\mu (x,y), \qquad \bar{y}=y+\epsilon_4(x,y),
\label{28} \\
\tilde{\epsilon}_4(p,y)&=&{8\pi GL\over p^2 + p/L}\tilde{T}(p)\exp(-py),
\qquad \tilde{\epsilon}_\mu(p,y)=-i {p_\mu\over p}\tilde{\epsilon}_4(p,y).
\eea{29}
The new coordinate system still obeys $\bar{h}_{\mu 4} =0$
everywhere in $\Sigma$, since $\epsilon_\mu$ obeys
\beq
\partial_\mu \epsilon_4 + \dot{\epsilon}_\mu=0.
\eeq{29a} 
The metric fluctuation in the new coordinate system is given by 
Eq.~(\ref{27}); it is finite in the limit $L\rightarrow \infty$.
Moreover, $\bar{h}_{\mu\nu}$ is linear in the source $T_{\mu\nu}$, and
small everywhere in $\Sigma$, if the energy density of the source is well
below the black-hole limit. This is in prefect analogy with 
standard Einstein's gravity.
 
The story does not end here, though, as it can be seen by closer 
inspection of Eqs.~(\ref{28},\ref{29}). 

Consider for simplicity a static, 
spherically-symmetric distribution of matter on the
brane, with total mass $M$. Outside matter, in the region $GM \ll 
r\equiv|\vec{x}| \ll L$,
the position of the brane in the new coordinate system is~\footnote{To linear
order $r = |\vec{\bar{x}}|$.} 
\beq
\epsilon_4(r,0)=  {2GM L\over r}.
\eeq{30}
This function can be large even in the region $r \gg GM$, so that the 
limit of validity of the linear approximation must be re-examined. 
Recall that
the metric transforms as follows under the reparametrization
$\bar{x}^\mu=x^\mu + \epsilon^\mu (x,y)$, $\bar{y}=y+\epsilon_4(x,y)$:
\beq
g_{\mu\nu}(x,y)= {\partial \bar{z}^a \over \partial x^\mu}
{\partial \bar{z}^b \over \partial x^\nu}\bar{g}_{ab}(\bar{z}),
\qquad \bar{z}^a=\bar{x}^\mu,\bar{y}.
\eeq{31}
If we expand to linear order in $\bar{h}_{\mu\nu}$, 
and to {\em quadratic} order in $\epsilon_a$, we find, at $y=0$,
\beq
h_{\mu\nu}(x,0)=\bar{h}_{\mu\nu}(x) + \partial_\mu \epsilon_\nu (x,0) + 
\partial_\nu \epsilon_\mu (x,0) +  \partial_\mu \epsilon^\alpha (x,0) 
 \partial_\nu \epsilon_\alpha (x,0) + \partial_\mu \epsilon^4 (x,0) 
 \partial_\nu \epsilon_4 (x,0).
\eeq{32}
Notice that the last term in this expansion is {\em not} a 4-d 
reparametization and that when $\epsilon_4$ is given by Eq.~(\ref{30}) it is
$O(G^2M^2L^2/r^4)$.

Clearly, when we assume that matter is so diluted that 
$|\bar{h}_{\mu\nu}|\ll 1$ everywhere on the brane, 
the linear approximation for 
$\bar{h}_{\mu\nu}$ is justified by assumption, but the 
linear approximation for $\epsilon_a$ breaks down when 
$\partial_\mu \epsilon^4\partial_\nu \epsilon_4 \gg \bar{h}_{\mu\nu}$. 
This happens when $G^2M^2L^2/r^4 \gg GM/r$, i.e. when
\beq
r^3 \ll GML^2.
\eeq{33}
This is exactly the condition found in ref.~\cite{nyu3}.

The breakdown of the linear approximation for $\epsilon_a$ means that, 
in the region $r \leq (GML^2)^{1/3}$, the 
position of the brane is still $\bar{y}=\epsilon^4(x,0)$, but 
$\epsilon^4(x,0)$ is no longer given by Eq.~(\ref{30}).  
To study the brane inside that region, we choose
$\epsilon^4$ and $\epsilon^\mu$ by demanding only that $g_{\mu5}=0$ and 
that $\epsilon^4$ is still given by Eq.~(\ref{30}) at large distances:
\beq
\epsilon_4(r,0) \approx {2GM L\over r}, \qquad \mbox{for } 
r\geq (GML^2)^{1/3}.
\eeq{34}
We can always set $\epsilon^\mu(x,0)=0$ with a 4-d coordinate transformation. 
The metric fluctuation is then
\beq
h_{\mu\nu}(x,0)=\bar{h}_{\mu\nu}(x) + 
\partial_\mu\epsilon^4(x,0)\partial_\nu\epsilon_4 (x,0).
\eeq{35}

To find the metric on the brane, we begin by making the following ansatz:
\beq
\bar{h}_{00}(r) = {2F(r)M\over  r}, \qquad 
\bar{h}_{ii}(r)= {F(r)M\over  r}.
\eeq{35a}
The asymptotic behavior ot the function $F(r)$ is:
\beq
F(r)=G \mbox{ for } r\ll (GML^2)^{1/3}, \qquad
F(r)={4\over 3}G \mbox{ for } r\gg (GML^2)^{1/3};
\eeq{35b}
otherwise, $F(r)$ is arbitrary.

The linearized scalar curvature of the ansatz vanishes identically everywhere. 
At large distances, $r\gg (GML^2)^{1/3}$, it
approximates the metric of the linearized DGP equations~\cite{nyu1,nyu3,gr}
[see also Eq.~(\ref{27})].

Suppose now that a 5-d diffeomorphism $\epsilon_4 $ exists, such that 
a) it obeys Eq.~(\ref{34}), b) the metric $h_{\mu\nu}$, given by 
Eq.~(\ref{35}),
solves the linearized 4-d Einstein equations in the region 
$r\ll (GML^2)^{1/3}$.
With these assumptions, we can write the solution to Eq.~(\ref{22}) as
\beq
g_{\mu\nu}(x,0)=\eta_{\mu\nu} + h_{\mu\nu}(x,0) + \Delta_{\mu\nu}(x,0),
\eeq{37}
with $|\Delta_{\mu\nu}(x,0)|\ll |h_{\mu\nu}(x,0)|$ {\em everywhere} on the 
brane~\footnote{We have assumed again that matter is diluted, i.e. that 
$|h_{\mu\nu}|\ll 1$ everywhere on the brane. This assumption has been made
for clarity's sake and can be relaxed.}.
The last statement can be proven by approximating Eq.~(\ref{22}) as
\bea
L_{\mu\nu,\rho\sigma}\Delta^{\rho\sigma}&=& J_{\mu\nu}, \label{38}\\
J_{\mu\nu} &=& {2\over L} K_{\mu\nu} + 16\pi GT_{\mu\nu} - 
L_{\mu\nu,\rho\sigma}h^{\rho\sigma}.
\eea{39}
The source $J_{\mu\nu}$ is conserved, and everywhere smaller
than $L_{\mu\nu,\rho\sigma}h^{\rho\sigma}$, because $h_{\mu\nu}$ solves  
by assumption the Einstein equations, for $r^3\ll GML^2$, and, for 
$r^3\gg GML^2$, it solves by construction the linearized DGP equations. 

Conservation of the source ensures that Eq.~(\ref{38}) can be solved, while
$|J_{\mu\nu}|\ll |L_{\mu\nu,\rho\sigma}h^{\rho\sigma}|$ guarantees 
that $|\Delta_{\mu\nu}(x,0)| \ll |h_{\mu\nu}(x,0)|$. 
Extending $h_{\mu\nu}$ and $\Delta_{\mu\nu}$ to the interior of $\Sigma$ is
straightforward because $h_{\mu\nu}+\Delta_{\mu\nu}$ obeys Eq.~(\ref{25}).

At this point, we are left only with the task of finding a shift $\epsilon^4$
which satisfies our assumption.
In the spherically symmetric case, we notice that the Schwarzshild metric
at $r\gg GM$ is $h_{00}(r)= 2G M/r$, $h_{ii}=2G M/r$, $i=1,2,3$, so that
the diffeomorphism we need is
\beq
\epsilon_4(r,0)= 2\sqrt{G Mr}.
\eeq{41}

Let us conclude with a few remarks.
\begin{itemize}
\item The limit of validity of Eq.~(\ref{41}) can be found by demanding that
the contribution to the extrinsic curvature due to the brane bending in 
Eq.~(\ref{41}) is smaller than that given by Eq.~(\ref{30}). Since
the curvature due to bending is $\sim |d^2\epsilon_4/dr^2|$ we find 
$r\ll (GML^2)^{1/3}$. Therefore, the domain of validity of Eq.~(\ref{41})
is complementary to that of  Eq.~(\ref{30}). 
\item The fact that quadratic corrections to the linear approximation cure the
vDVZ discontinuity is at the hart of Refs.~\cite{v,nyu2,nyu3}. In this paper, 
we spelled out that it is the linear approximation 
{\em for the fluctuations of the brane} that fails at $r\ll (GML^2)^{1/3}$, 
{\em not the linearization of the 5-d metric} (see also~\cite{nyu2}). 
\item  The previous observation makes the breakdown of linearity at such large
distances more palatable, since the brane is almost tensionless, and can, 
therefore, bend significantly even over macroscopic (astronomical) distances.
\item When the position of the brane is given by Eq.~(\ref{41}), 
the sub-leading correction to the induced metric, $\Delta_{\mu\nu}$, is
proportional to $\sqrt{GM}$ [see Eqs.~(\ref{38},\ref{39})]. It is tantalizing 
to conjecture that this correction may give rise to interesting modifications 
of Newtonian dynamics at some macroscopic length scale.
\item Absence of a vDVZ discontinuity is only 
a qualified good news for the DGP theory. 
Indeed, the very breakdown of the linear approximation at the macroscopic
lenght scale $r=(GML^2)^{1/3}$ may signal 
that the 4-d scalar $\epsilon_4(x,0)$ 
interacts strongly with the stress-energy tensor at the quadratic level --for
instance through an interaction term $\sim L\sqrt{G}(\epsilon_4)^2 T$.  
\end{itemize}
\subsection*{Acknowledgments}
We should like to thank G. Dvali, A. Lue, and C. Deffayet for interesting 
comments. 
This work is supported in part by NSF grant PHY-0070787.

\end{document}